\documentclass[11pt]{JHEP3}
\usepackage{amsmath}
\usepackage{amssymb}
\usepackage{bbold}
\usepackage{xspace}
\usepackage[numbers]{natbib}
\usepackage[bottom]{footmisc}
\usepackage{graphicx}
\usepackage{psfrag}

\DeclareMathOperator{\Tr}{Tr}

\DeclareMathOperator{\dn}{dn}
\DeclareMathOperator{\sn}{sn}
\DeclareMathOperator{\cn}{cn}
\newcommand{\eqn}[1]{(\ref{#1})}  
\newcommand{\om}{\sqrt{\omega_2^2-\omega_1^2}} 
\newcommand{\adss}{$\text{AdS}_5\times S^5$\xspace}

\title{Splitting spinning strings in AdS/CFT}
\author{Kasper Peeters, Jan Plefka and Marija Zamaklar\\
 Max-Planck-Institut f\"ur Gravitationsphysik\\
Albert-Einstein-Institut\\
Am M\"uhlenberg 1\\ D-14476 Golm, Germany\\

\email{kasper.peeters@aei.mpg.de, jan.plefka@aei.mpg.de, marija.zamaklar@aei.mpg.de}}
\preprint{AEI-2004-094\\hep-th/0410275}

\abstract{We study the semiclassical decay of macroscopic spinning
strings in \adss through spontaneous splitting of the folded string
worldsheet. Based on similar considerations in flat space this decay
channel is expected to dominate the full quantum computation.
The outgoing strings are uniquely specified by an infinite
set of conserved (local) charges with a regular expansion in inverse
powers of the initial angular momentum.  We compute these charges and determine
functional relations between them.  Finally, a preliminary discussion of the
corresponding calculation in the non-planar sector of the dual gauge theory
is presented.}

\begin{document}
\section{Introduction and summary}

Within the framework of the AdS/CFT correspondence it has proved
fruitful to explore sectors of both theories which are characterized
by large charges. In the Berenstein, Maldacena and Nastase
limit~\cite{Berenstein:2002jq} one considers operators of U(1)$_R$
charge~$J$ such that $J\sim N^2$ as $N\rightarrow\infty$ with $g_{\rm
YM}$ finite.  This leads to a subsector of the theory for which the
quantum corrections are under control, despite the fact that the
't~Hooft coupling $g^2_{\rm YM}\, N$ becomes large. A remarkable
feature of this limit is that it maintains a full genus expansion
\cite{Constable:2002hw,Kristjansen:2002bb}.  The effective genus
parameter turns out to be $g_2=J^2/N$, which is nonvanishing despite
the \mbox{large-$N$} limit.  This allows one to compare the process of
string splitting with a computation in the dual gauge theory, by
determining decay widths of the corresponding BMN
operators~\cite{Freedman:2003bh}. At leading order in $\lambda'=g_{\rm
YM}^2 N/J^2$ and $g_2$ these computations have been shown to agree
\cite{Gutjahr:2004qj} with light-cone string field
theory~\cite{Spradlin:2002ar}.

Following the BMN idea, one can consider subsectors of the gauge
theory with several large 
charges~\cite{Gubser:2002tv,Frolov:2003qc,Frolov:2003xy,Beisert:2003xu}.  
As before,
the operators in these sectors have controlled quantum corrections,
allowing one to make a direct comparison with string theory. The objects
dual to such operators have been identified as large, macroscopic
spinning strings in \adss \cite{Frolov:2003qc,Frolov:2003xy,Arutyunov:2003uj}. 
The energies of these strings and the
anomalous dimensions of the gauge theory operators are in agreement up
to two-loop order in $\lambda'$~\cite{Beisert:2003tq,Beisert:2003xu,Beisert:2003ea,Serban:2004jf}
\footnote{They actually start
to disagree at the three loop level \cite{Beisert:2003ys}. This mismatch
has been argued to arise from the non-commutativity of taking $J\to \infty$ and
expanding in $\lambda'$ \cite{Beisert:2004hm}.}.
Clearly however, these computations only probe zeroth-order effects in the string
coupling constant~$g_s$.

The central question to be addressed in this work is what can be said
about \mbox{$g_s \neq 0$} effects for large spinning
strings.  In order to study string splitting, one would
in principle need to compute the decay widths of long spinning strings
and compare these to a dual computation of gauge theory operators with
many impurities. Unfortunately, the determination of decay widths in
quantum \adss string theory  is at least for the
time being out of reach.

\begin{figure}[t]
\begin{center}
\hbox{\includegraphics*[width=.4\textwidth, angle=-90]{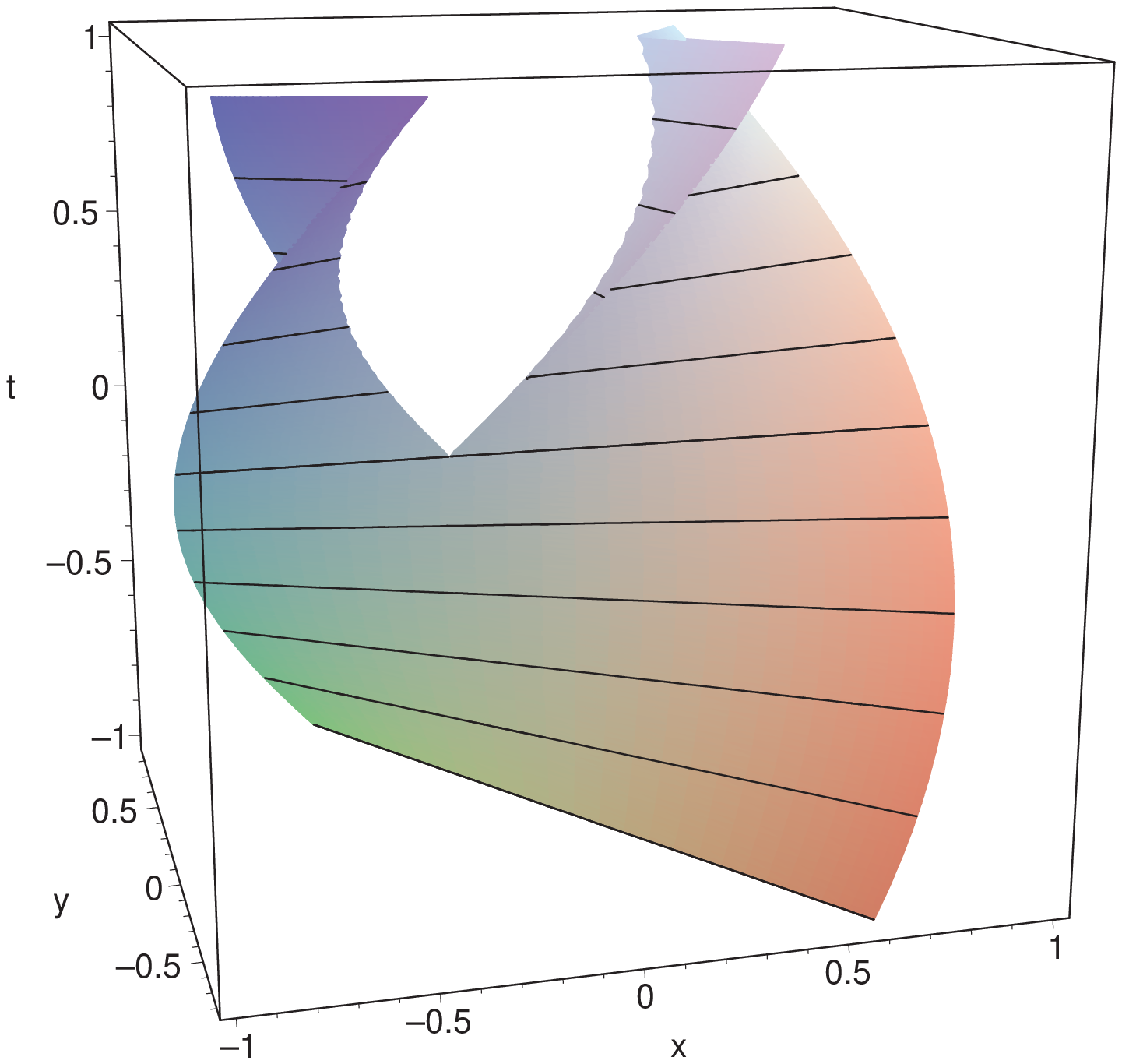}\hskip-4em
      \includegraphics*[width=.4\textwidth, angle=-90]{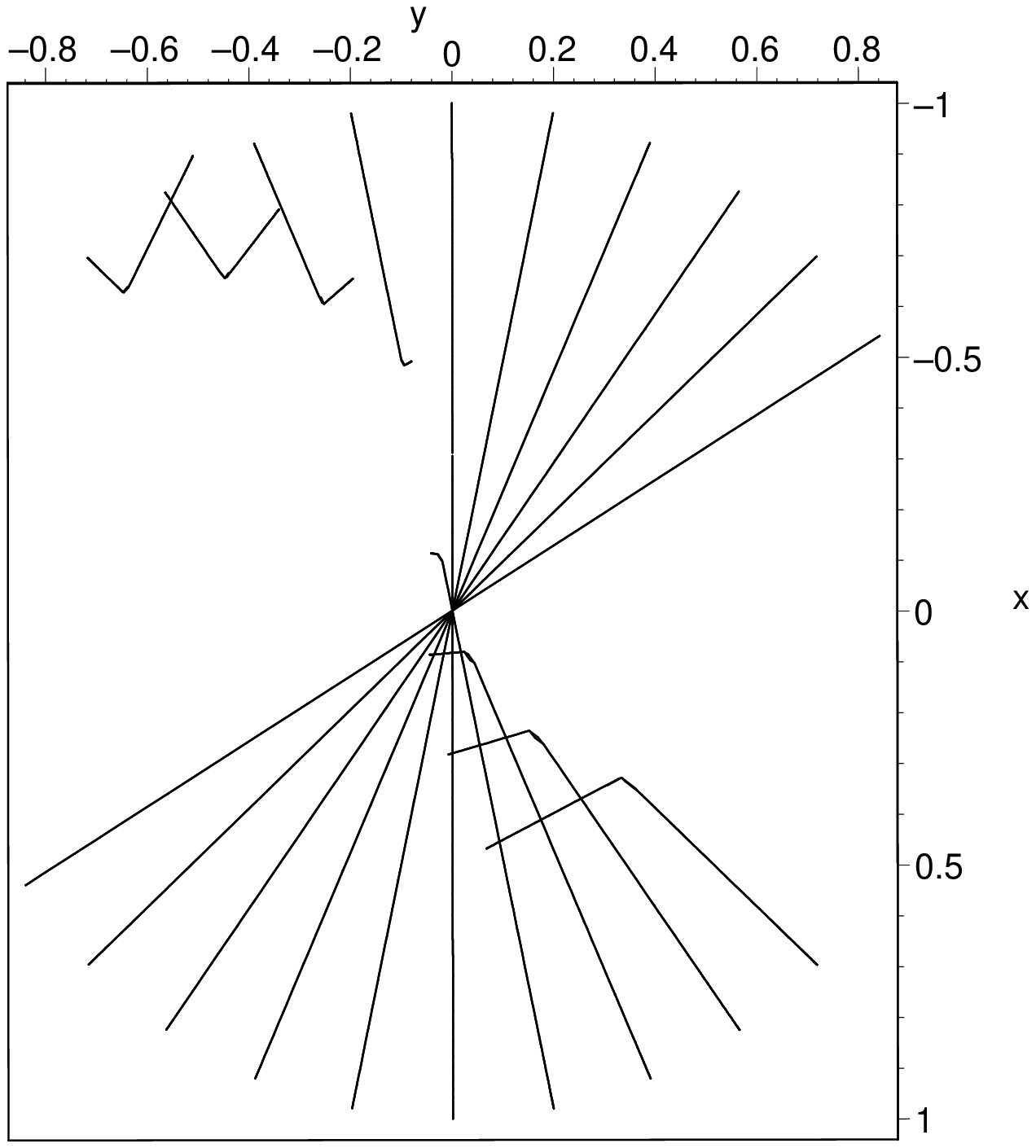}}
\end{center}
\caption{Semi-classical decay of a folded, rotating string in flat
  space-time, following~\cite{Iengo:2003ct}. The plot on the right
  shows snapshots at various values of $\tau$. The outgoing pieces
  exhibit kinks, which propagate outward along the strings. New
  momenta $P_x^I= -P_x^{II}$ are generated in the decay process.}
\label{f:splitflat}
\end{figure}

It is, however, possible to analyze the decay semi-classically. In
\emph{flat} space-time, the semi-classical decay of macroscopic
strings was analyzed in detail by Iengo and
Russo~\cite{Iengo:2003ct}. They also compared the
semi-classical results to those of a full quantum treatment. In the
semi-classical approach, one starts with a classical, rotating closed
string solution. At a given time~$\tau=0$, the string can
spontaneously split if two points~$\sigma$ and $\sigma'$ on the string
coincide in target space, and if their velocities agree. The string
described by these boundary
conditions,~$X^\mu(\tau,\sigma)=X^\mu(\tau,\sigma')$ and $\dot
X^\mu(\tau,\sigma)=\dot X^\mu(\tau,\sigma')$, then forms a ``figure
eight''. The splitting is realized by declaring that from $\tau=0$
onward, each of the two string pieces (``left and right'' from the
overlapping point), \emph{separately} satisfy periodic boundary
conditions. The initial conditions on the positions and velocities of
the outgoing pieces are simply taken to be those of the incoming
string at the moment~$\tau=0$ of splitting.  The effect of the
splitting propagates with the speed of light along the outgoing
pieces, leading to kink-like shapes (see figure~\ref{f:splitflat}).

\begin{figure}[t]
\begin{center}
\includegraphics*[width=0.5\textwidth]{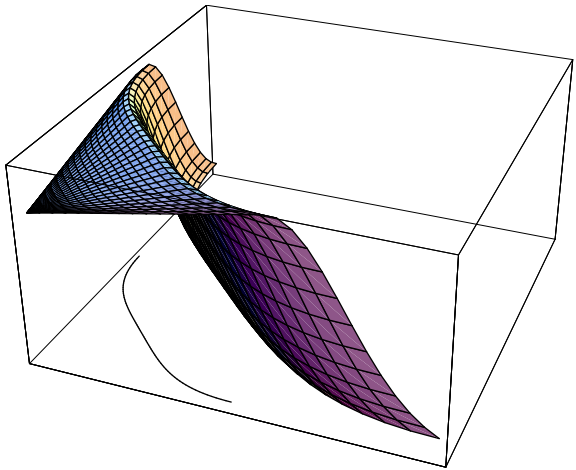}
\caption{Sketch of the relation between the semi-classical and the
  full quantum calculations. The surface depicts the quantum decay
  amplitude over the (horizontal) plane spanned by the mass-square of
  the two outgoing strings, $(M_I)^2$ and $(M_{II})^2$. The amplitude
  reaches its maximum over the curve allowed by semi-classical decay.}
\label{f:classquant}
\end{center}
\end{figure}

The relations between the energies and angular momenta of the outgoing
strings are determined completely by conservation laws, i.e.~one does
not need to derive the explicit string shapes in order to obtain these
relations. From the relations between the charges one can then produce
a curve in, for instance, the plane spanned by the masses $M_I$ and
$M_{II}$ of the outgoing string pieces. In flat space-time, this curve
can be compared with a \emph{full quantum} string computation of the
decay rate. It has been shown~\cite{Chialva:2003hg,Iengo:2002tf} that
the quantum decay rate, as a function of the outgoing masses, reaches
its maximum very close to the curve obtained from the classical
analysis (see figure~\ref{f:classquant}).  In order to understand this
relation between the semi-classical computation and the full quantum
treatment, it is important to realize that the space of kinematically
allowed decays of a string is much larger than the channel which is
available using a semi-classical treatment (in the sense described
above). Hence, the semi-classical analysis only describes one
particular decay channel. Fortunately the quantum analysis
of~\cite{Chialva:2003hg,Iengo:2002tf} shows that this is the most
probable channel (at least in flat space-time).

\medskip

In the present paper we will analyze the decay of semi-classical
strings on \adss, with the goal of producing predictions which can in
principle be verified on the gauge theory side. We will focus on the
folded string which is rotating on the~$S^5$ factor of the
background. We first review the properties of this folded spinning
string and all its charges. There are ``global'' charges (angular
momenta and energy), associated to the isometries of the target space,
as well as an infinite set of ``local'' (commuting) charges, related
to the integrable structure of the string sigma model. Due to the fact
that we consider a rigidly rotating string, all local charges are
uniquely determined as functions of the global charges.

We then analyze the semi-classical splitting process. This splitting
process introduces one new parameter $a$, the splitting parameter, and
all the local and global charges of the outgoing strings are uniquely
determined as functions of this parameter and the global charges of
the incoming string. This is now the case, despite the fact that the
outgoing strings are highly \emph{non-rigid},
i.e.~strongly~\emph{fluctuating} (as in the flat-space case depicted
in figure~\ref{f:splitflat}).  It is a consequence of the
``continuous'' way in which the worldsheets of in- and out- strings
have been glued together.  Since both local and global charges are
conserved under the free ($g_s=0$) evolution of the string, they can
be computed by integrating the charge densities at the moment of
splitting, i.e.~at the moment when the charge densities of the in- and
out-going strings coincide. In this way, one circumvents the need to
explicitly construct the solutions for the outgoing strings.

In the decay process, several charges which were zero for the ingoing
string now get turned on for the individual outgoing pieces. This is
similar to the decay in flat space-time, where one generates new
momenta $P_x^I=-P_x^{II}$ (the pieces move away from each other). For
the string on~$S^5$, new angular momentum components get turned on.
These new angular momenta are neither Casimirs nor elements of the
Cartan subalgebra, indicating that the outgoing strings are
descendants rather than highest-weight states in the gauge theory
dual.  Having found these new charges, we in addition also construct
the generating functional for all the higher charges as a function of
the charges of the incoming string and the splitting parameter~$a$.

The conservation of an infinite set of local charges is a consequence
of our construction. It can be checked explicitly that the
\emph{quantum} decay of the string in flat space does not preserve the
higher charges.  However, for the most dominant, semiclassical decay
processes, the conservation laws do hold for all charges. This
suggests that the conjectured integrability in the planar sector
of~${\cal N}\!=\!4$ super-Yang-Mills may be extended in a certain
sense to the non-planar sector.
\medskip

In order to make a comparison to the gauge theory, it is necessary to
express the decay process on the string theory side purely in terms of
relations between charges. In principle there is an infinite set of
relations, for all the local charges, but we focus on the relations
between the global charges. We present these relations in
section~\ref{s:relations}. The gauge theory quantum amplitude is
expected to attain a maximum over the semi-classical string decay
curves, just as in figure~\ref{f:classquant}.

On the gauge theory side, the decay of long strings can be analyzed in
the spin chain picture of~\cite{Minahan:2002ve}. The splitting
operator, which when acting on a single-trace operator produces a
double-trace operator, is represented by the non-planar part of the
dilatation operator. In section~\ref{s:gauge_theory} we discuss these
gauge theory ingredients. We argue that, despite the fact that we are
not in the BMN regime, an effective genus counting parameter~${\cal
J}^2/N$ is still present. We also discuss the higher conserved
charges and their role in reducing the possible spin chain decay
channels. A detailed study of this decay is, as we will argue,
hampered by the complexity of the spin chain wave functions, and we
leave this for future work.

Finally, let us note that our construction can easily be extended to
the study of strings in less symmetric backgrounds (and in particular
to backgrounds dual to confining gauge theories). This could provide one
with information about meson and glue-ball decays. A study of these
processes is under investigation.

\newpage

\section{Review of the folded string solution}

In order to set up our notation and introduce the new charges that
will be generated in the process of the decay, let us briefly review
the construction of the folded two spin string rotating in $S^5$, as
first presented by Frolov and Tseytlin~\cite{Frolov:2003xy}. Using the
parameterization
\begin{equation}
\label{e:para}
X_1 + i X_2 = \sin \gamma\,  \cos \psi \,e^{i{\varphi}_1} \, , \quad X_3 + i X_4 =
\sin \gamma\, \sin \psi \, e^{i{\varphi}_2} \, , \quad X_5 + i X_6 = \cos \gamma\,
e^{i{\varphi}_3} \, ,
\end{equation}
the metric on a five-sphere $X_1^2 + \cdots + X_6^2 = 1$ can be written as
\begin{equation}
{\rm d}s_{S^5}^2 = {\rm d}\gamma^2  + \cos^2 \gamma\, {\rm d} {\varphi}_3^2 +
\sin^2 \gamma ( {\rm d} \psi^2 + \cos^2 \psi\,  {\rm d} {\varphi}_1^2 +
\sin^2 \psi\,  {\rm d} {\varphi}_2^2 ) \, ,
\end{equation}
whereas the metric on $\text{AdS}_5$ reads
$
{\rm d}s_{\text{AdS}_5}^2 = {\rm d}\rho^2  - \cosh^2 \rho\, {\rm d} t^2 +
\sinh^2 \rho \, {\rm d} \Omega_3^2 
$.
The two-spin string solution is given by the equations
\begin{equation}
\label{e:FT2}
t = \kappa \tau \, , \quad \rho =0 \, , \quad \gamma = \frac{\pi}{2} \, , 
\quad \varphi_3 =0 \, , \quad \varphi_1 = w_1 \tau \, , \quad \varphi_2 = w_2 
\tau \,, \quad \psi = \psi(\sigma) \, , 
\end{equation}
where $\kappa,w_1$ and $w_2$ are constants. The equation which
determines the profile of $\psi(\sigma)$ is
\begin{equation}
\label{e:eomi}
\psi^{''} + {1\over 2}  w_{21}^2 \sin(2\psi) = 0 \, , \quad w_{21}^2 \equiv w_2^2 
- w_1^2 \geq 0 \, . 
\end{equation}
By integrating this equation once, we obtain the following equation,
\begin{equation}
\label{e:eoms}
\psi^{'2}  = w_{21}^2 (\sin^2 \psi_0 - \sin^2 \psi) \, .
\end{equation}
Here the constant $\psi_0$ corresponds to the target-space length of
the folded string; the point at which the first derivative of $\psi$
vanishes is the point at which the world-sheet of the string turns
back onto itself.  The conformal gauge constraints imply
\begin{equation}
\label{conf}
\kappa^2 = w_2^2 \sin^2 \psi_0 + w_1^2 \cos^2 \psi_0 \, .
\end{equation}
The motion of the string is confined to a three-sphere embedded in
the five sphere, which will remain true also for the two outgoing
strings  after the decay process to be  considered in the next 
section. As the isometry group of the
three sphere is $SO(4)$, there are 6~conserved angular momenta $J_{ij}$, 
associated to the 6 Killing vectors,
\begin{equation}
J_{ij} = \sqrt{\lambda} \int_0^{2\pi} {{\rm d} \sigma \over 2 \pi} (X_i \dot{X}_j 
-  X_j \dot{X}_i) \equiv \sqrt{\lambda} {\cal J}_{ij} \, , \quad (i,j=1\cdots 4) \, .
\end{equation}
Explicitly, using the parameterization (\ref{e:para}) these can be
rewritten as (using $\gamma = {\pi \over 2}$)
\begin{align}
\label{e:J12}
{\cal J}_{12}  &= \int_0^{2 \pi} {{\rm d} \sigma \over 2 \pi}  \cos^2 \psi  
\dot{\varphi}_1\,, \\
\label{e:J34}
{\cal J}_{34} &=  \int_0^{2 \pi} {{\rm d} \sigma \over 2 \pi}  \sin^2 \psi  \dot{\varphi}_2\,, \\
\label{e:J13}
{\cal J}_{13} &=  \int_0^{2 \pi} {{\rm d} \sigma \over 2 \pi} \big( \cos {\varphi}_1 \cos{\varphi}_2 \dot{\psi} + \sin \psi \cos\psi (- \sin \varphi_2 \cos \varphi_2 \dot{\varphi}_2 + \sin \varphi_1 \cos \varphi_2 \dot{\varphi}_1) \big) \,, \\
\label{e:J24}
{\cal J}_{24} &=  \int_0^{2 \pi} {{\rm d} \sigma \over 2 \pi} \big( \sin {\varphi}_1 \sin{\varphi}_2 \dot{\psi} + \sin \psi \cos\psi ( \sin \varphi_1 \cos \varphi_2 \dot{\varphi}_2 - \sin \varphi_2 \cos \varphi_1 \dot{\varphi}_1)  \big)\,, \\
\label{e:J14}
{\cal J}_{14} &=  \int_0^{2 \pi} {{\rm d} \sigma \over 2 \pi} \big( \cos {\varphi}_1 \sin{\varphi}_2 \dot{\psi} + \sin \psi \cos \psi ( \cos \varphi_1 \cos \varphi_2 \dot{\varphi}_2 + \sin \varphi_2 \sin \varphi_1 \dot{\varphi}_1) \big) \,,  \\
\label{e:J23}
{\cal J}_{23} &=  \int_0^{2 \pi} {{\rm d} \sigma \over 2 \pi} \big( \sin{\varphi}_1 \cos \varphi_2 \dot{\psi} - \sin \psi \cos \psi ( \sin \varphi_1 \sin \varphi_2 \dot{\varphi}_2 + \cos \varphi_2 \cos \varphi_1 \dot{\varphi}_1) \big) \,.  
\end{align}
An additional charge is the energy~$E$, which is associated to the
translation invariance with respect to global time. Using the
constraint~\eqn{conf} one finds that the energy is given by
\begin{equation}
\label{e:En} 
E = \sqrt{\lambda} \int_{0}^{2 \pi}  {{\rm d} \sigma \over 2 \pi} \dot{X}^0 =
 \sqrt{\lambda}\,  \kappa = \sqrt{\lambda}\,  \sqrt{w_2^2 \sin^2 \psi_0 + w_1^2 
\cos^2 \psi_0} \, .
\end{equation}
Before the decay, the string (\ref{e:FT2}) carries two (mutually commuting)
angular momenta, $J_{12}$ and $J_{34}$
\begin{align}
\label{e:J12b}
{\cal J}_{12} &= {2 w_1 \over \pi w_{21}} \int_0^{\psi_0} {\cos^2 \psi\,
  {\rm d} \psi \over \sqrt{\sin^2 \psi_0 - \sin^2 \psi}} =
\frac{2\omega_1}{\pi\, \omega_{21}} E(q) \, , \\
\label{e:J34b}
{\cal J}_{34} &= {2 w_2 \over \pi w_{21}} \int_0^{\psi_0} {\sin^2 \psi\,
  {\rm d} \psi \over \sqrt{\sin^2 \psi_0 - \sin^2 \psi}} =
\frac{2\omega_2}{\pi\, \omega_{21}} (K(q)-E(q)) \, , \quad q\equiv \sin^2 \psi_0\,.
\end{align}
The expressions on the right-hand sides have been obtained by making a
change of variables to $\psi'$, defined by $\sin\psi/
\sin\psi_0=\sin\psi'$. \footnote{Our conventions for the elliptic
integrals are
\begin{equation}
\label{eq:conven}
E(x;q)=\int_0^x\!{\rm d} \varphi\, \sqrt{1-q\, \sin^2\varphi}\,, \qquad\qquad
F(x;q)=\int_0^x\!{\rm d} \varphi \frac{1}{\sqrt{1-q\, \sin^2\varphi}} \,.
\end{equation}
with $E(q):=E(\pi/2;q)$ and $K(q):=F(\pi/2;q)$.}

Although the worldsheet densities for the other angular momenta are
non-zero before the split, they vanish when integrated over the
world-sheet. The two non-vanishing angular momenta are related by
\begin{equation}
\label{e:J12E}
1=\frac{{\cal J}_{12}}{\omega_1}+ \frac{{\cal J}_{34}}{\omega_2}\,.
\end{equation}
as a consequence of~(\ref{e:J12}) and~(\ref{e:J34}).
Moreover one derives from (\ref{e:J12b}), (\ref{e:J34b}) and (\ref{e:En})
\begin{equation}
\label{e:omK}
\sqrt{\omega_2^2-\omega_1^2}=\frac{2}{\pi}\, K(q)\,, \qquad q=
\frac{\kappa^2-\omega_1^2}{\omega_2^2-\omega_1^2} \, .
\end{equation}
It then follows that all information to determine the energy E=$\sqrt{\lambda}\,
{\cal E}({\cal J}_{12},{\cal J}_{34})$ as a function of the angular momenta
lies within the two equations
\begin{align}
\label{e:qeq}
\frac{4}{\pi^2}\, q &= \frac{{\cal E}^2}{K(q)^2} - \frac{{\cal J}_{12}^2}{E(q)^2}\,, \\
\label{e:Eeq}
\frac{4}{\pi^2} &= \frac{{\cal J}_{34}^2}{(K(q)-E(q))^2} - 
\frac{{\cal J}_{12}^2}{E(q)^2}\,.
\end{align}
upon elimination of $q$. This may be achieved iteratively in an expansion 
for large total angular momentum
${\cal J}={\cal J}_{12}+{\cal J}_{34}$ via the ansatz
\begin{eqnarray}
\label{qexp}
q&=& q_0 + \frac{q_1}{{\cal J}^2} + \frac{q_2}{{\cal J}^4} + \ldots\,, \\
\label{Enexp}
{\cal E}&=& {\cal J}\, {\cal E}_0 + \frac{{\cal E}_1}{{\cal J}} + \frac{{\cal E}_2}
{{\cal J}^3} + \ldots\,.
\end{eqnarray}
One finds that $q_0$ is determined by the equation
\begin{equation}
\label{e:q0}
\frac{E(q_0)}{K(q_0)} = 1- \alpha\,, \qquad\qquad \alpha:= \frac{{\cal J}_{34}}{{\cal J}}\,.
\end{equation}
All the higher terms $q_i$ with $i>0$ are then given algebraically as
functions of~$q_0$, and similarly for~${\cal E}_i$. The first
non-trivial terms which one finds are
\begin{equation}
\label{x1def}
q_1=-\frac{4\, E(q_0)\, K(q_0)^2\, (K(q_0)-E(q_0))\, (1-q_0)\, q_0}{\pi^2
\, (E(q_0)^2-2\, E(q_0)\, K(q_0)\, (1-q_0)+(1-q_0)\, K(q_0)^2)} \,,
\end{equation}
as well as
\begin{equation}
\label{e:E0E1}
{\cal E}_0=1\,, \quad
{\cal E}_1= \frac{2}{\pi^2}\, K(q_0)\, (E(q_0)-(1-q_0)\, K(q_0))
\,.
\end{equation}

\section{Semiclassical decay of the folded string}

\subsection{The splitting}
Let us now consider the spontaneous splitting of the
solution~\eqn{e:FT2}.  As mentioned in the introduction, we will focus
on the relations satisfied by the charges of the outgoing strings. In
order to obtain those relations, we fortunately do not need to derive
the precise form of the solutions $X^I_\mu(\tau,\sigma)$ for the
outgoing strings.  While those solutions could be obtained in flat
space-time~\cite{Iengo:2003ct}, it would be much more difficult to do
so in the \adss background.

We choose a parametrization on the world-sheet such that
at the ``end''-point of the folded, unsplit string we have
$\psi(\sigma=0)=-\psi_0$ on one end of the string and
$\psi(\sigma=\pi)=\psi_0$ on the other end. The splitting occurs on
the worldsheet at points~$\sigma=\pi a$ and~$\sigma=-\pi a$, which are
both mapped to the same target space point $\psi(- \pi a) = \psi(\pi
a)= \tilde {\psi}$. The setup is thus

\begin{center}
\psfrag{ps0}{\small $\psi(0)=-\psi_0$}
\psfrag{ps2p}{\small $\psi(2\pi)$}
\psfrag{psap}{\small $\psi(a\pi)=\tilde\psi$}
\psfrag{ps-ap}{\small $\psi(-a\pi)$}
\psfrag{psp2}{\small $\psi(\pi)=\psi_0$}
\psfrag{cut}{\small cut}
\includegraphics*[width=.6\textwidth]{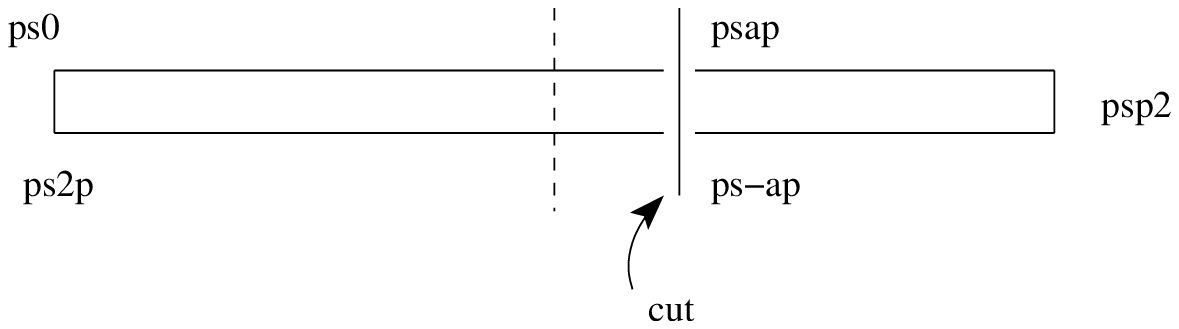}
\end{center}

The charges carried by the outgoing strings can be calculated by
evaluating the expressions~\eqn{e:J12}-\eqn{e:En}, using the solution
of the string \emph{before} the decay, but integrating them over the
lengths of each piece of string separately (i.e.~$\sigma\in [-\pi a, \pi a]$
for the first piece and $\sigma \in [-\pi,-\pi a] \cup [\pi a,\pi]$ for the
second one). In these calculations the string solution is evaluated at
the moment of the decay, which without loss of generality we will take
to be~$\tau_{d} =0$. This is consistent, as the initial conditions generated
from the unsplit solution for the outgoing two string pieces are consistent, i.e.
obey the Virasoro constraint.

The splitting parameter~$a$ is then related to the splitting
point~$\tilde{\psi}$ via (\ref{e:eoms}) by
\begin{equation}
\label{e:K+F}
2\pi a = \frac{2}{\omega_{21}}\int_{-\psi_0}^{\tilde\psi}
\frac{{\rm d}\psi}{\sqrt{\sin^2\psi_0 -\sin^2\psi}}
= \frac{2}{\omega_{21}} \, \left ( K(q) + F(x;q) \right )
\end{equation}
where $x:=\arcsin(\frac{\sin\tilde\psi}{\sin\psi_0})$. The ``mirror''
equation to this is
\begin{equation}
\pi (1-a) \om = K(q) -F(x;q) \,.
\end{equation}
The angular momenta $J_{12}$ and $J_{34}$, which were non-zero before
the split, get distributed between the outgoing string pieces $I$ and
$II$ according to
\begin{align}
\label{e:J12I}
{\cal J}_{12}^I &= \frac{\omega_1}{\pi\omega_{21}}\left ( E(q) + E(x;q) \right )\,,
\\
\label{e:J34I}
{\cal J}_{34}^I &= \frac{\omega_2}{\pi\omega_{21}}\left ( K(q)-E(q) + F(x;q) - E(x;q) \right )\,, \\
\label{e:J12II}
{\cal J}_{12}^{II} &= \frac{\omega_1}{\pi\omega_{21}}\left ( E(q) - E(x;q) \right )\,, 
\\
\label{e:J34II}
{\cal J}_{34}^{II} &= \frac{\omega_2}{\pi\omega_{21}}\left (K(q) - E(q) -
F(x;q) + E(x;q) \right )\, .
\end{align}
Moreover one has as a consequence of the above
\begin{equation}
a = \frac{{\cal J}_{12}^I}{\omega_1} +\frac{{\cal J}_{34}^I}{\omega_2}\,, 
\qquad\qquad\qquad
1-a = \frac{{\cal J}_{12}^{II}}{\omega_1} +\frac{{\cal J}_{34}^{II}}{\omega_2}\,. 
\label{e:a,1-a}
\end{equation}
The remaining angular momenta (\ref{e:J13})--(\ref{e:J23}) vanish
before the split, but they become non-zero for the outgoing strings,
\begin{align}
\label{e:Jzero}
{\cal J}_{13}^{I} &= {\cal J}_{13}^{II} = 0 \, , \quad  \quad \quad {\cal J}_{24}^{I} = {\cal J}_{24}^{II} = 0  \, , \\
\label{e:Jnew1}
{\cal J}_{14}^{I} & = - {\cal J}_{14}^{II} =
 - { w_2 \over \pi w_{21}} \sqrt{\sin^2 \psi_0 - \sin^2 \tilde{\psi}} \, , \\
\label{e:Jnew2}
{\cal J}_{23}^{I}  &=  -{\cal J}_{23}^{II} =
{ w_1 \over \pi w_{21}} \sqrt{\sin^2 \psi_0 - \sin^2 \tilde{\psi}} \, .
\end{align}
The sum of each of these momenta is zero in accordance with the
conservation laws.  The conformal gauge constraint (\ref{conf})
remains untouched,
\begin{equation}
\label{e:qsplit}
q=\frac{\kappa^2-\omega_1^2}{\omega_2^2-\omega_1^2} \, ,
\end{equation}
and the energies of two outgoing string pieces are given by
\begin{equation}
\label{e:enIII}
{\cal E}^I=\kappa\, a \,,\qquad  {\cal E}^{II}=\kappa\, (1-a) \, .
\end{equation}
A further relation is satisfied by the newly generated angular
momenta and the splitting parameter $x$,
\begin{equation}
\label{e:music}
({\cal J}_{14}^{I})^2 - ({\cal J}_{23}^{I})^2 = \frac{1}{\pi^2}\, q^2\cos^2x\,.
\end{equation}
This equation determines $x$ as a function of $q$ and $\Delta:= ({\cal
J}_{14}^{I})^2 - ({\cal J}_{23}^{I})^2$. 

\subsection{Relations between outgoing charges}
\label{s:relations}

The goal now is to eliminate the parameters $x$ and $q$ related to the
splitting point and initial string length,
and express all conserved charges in terms of a minimal set of
independent ones. Recall that before the split, the input data which
determine all the string charges are the total momentum ${\cal J}$ and
the filling fraction $\alpha = {\cal J}_{34}/{\cal J}$. The energy
$\cal{E}(\alpha,{\cal J})$ is expressed in terms of these two
parameters through~\eqn{Enexp}, which can be compared with the gauge
theory order by order in the $1/{\cal J}$ expansion.

The split introduces only one extra free parameter, namely the
point~$x$ at which the string splits, while the number of measurable
charges doubles: $\alpha^I$,$\alpha^{II}$, ${\cal J}^I$ and ${\cal
J}^{II}$. Hence after the split, the number of dependent quantities,
as well as the number of functional relations between them (which
should be compared to the gauge theory) is larger. Depending on which
quantities we want to relate, the choice for the set of independent
parameters might be different.

The first functional relation we want to establish is the relation
between the two angular momenta carried by the first part of the string,
\begin{equation}
\label{e:beta1234def}
\beta_{12}:= \frac{{\cal J}_{12}^{I}}{{\cal J}_{12}} \,,\qquad
\beta_{34}:= \frac{{\cal J}_{34}^{I}}{{\cal J}_{34}} \; .
\end{equation}
where the total momentum ${\cal J}$ is expressed as
\begin{equation}
{\cal J} := \underbrace{{\cal J}_{12}^I +{\cal J}_{12}^{II}}_{=:{\cal J}_{12}} + 
\underbrace{{\cal J}_{34}^I +{\cal J}_{34}^{II}}_{=:{\cal J}_{34}}  \, .
\end{equation}
Combining the equations (\ref{e:J12I}) and (\ref{e:J34I}) with equations
(\ref{e:J12b}) and (\ref{e:J34b})  one deduces that
\begin{align}
\label{eq:beta-exp1}
\beta_{12}  &=  {1\over 2} \bigg( 1 + {E(x;q) \over E(q)} \bigg) \,, \\
\label{eq:beta-exp2}
\beta_{34}  & = {1\over 2} \bigg(1 +{F(x;q) - E(x,q) \over K(q) - E(q)} \bigg) \, .
\end{align}
The parameter $q$ appearing in these equations is given as a series in
$1/{\cal J}$ with coefficients fixed by the data of the unsplit string
(see equations (\ref{qexp}), (\ref{e:q0}) and (\ref{x1def})). The splitting
point $x$ should now be eliminated by a combination of
global charges of the outgoing strings, which is at our disposal. 
We could choose either
$\beta_{12}$, $\beta_{34}$ or something like the string length fraction
$\frac{{\cal J}_{12}^I+{\cal J}_{34}^{I}}{{\cal J}}$ as the free
parameter of the splitting process. Any such choice will lead to
an expansion of $x$ in $1/{\cal J}^2$,
\begin{equation}
\label{e:xexp}
x =  x_0 + \frac{x_1}{{\cal J}^2} + \frac{x_2}{{\cal J}^4} + \ldots \, .
\end{equation}
In the remainder of this section we shall choose $\beta_{12}$ as the
new parameter. This yields the first two coefficients of \eqn{e:xexp}
as
\begin{align}
\beta_{12}  &=  {1\over 2} \bigg( 1 + {E(x_0;q_0) \over E(q_0)} \bigg)\,,\\
x_1 &= \frac{q_1}{2q_0} \frac{E(q_0) F(x_0; q_0) - K(q_0) E(x_0;
  q_0)}{E(q_0)\, \sqrt{1-q_0\,\sin^2 x_0}}\,,
\end{align}
where $q_1$ is given in~\eqn{x1def} and induces a $1/{\cal J}^2$ expansion for
$\beta_{34}$.  Substituting the expansion for
$q$ and $x$ in the second equation~\eqn{eq:beta-exp2}, one is left with
the desired first functional relation, namely $\beta_{34} = \beta_{34}(\beta_{12}, \alpha,
{\cal J})$, given as a series in $1/{\cal J}$
\begin{align}
\label{e:b34pert}
\beta_{34} &= \beta_{34}^0 + {\beta_{34}^1 \over {\cal J}^2} + \ldots \,, \\
\beta_{34}^0 & = {1\over 2} \bigg(1 +{F(x_0;q_0) - E(x_0,q_0) \over K(q_0) - E(q_0)} \bigg)\,,\\
\beta_{34}^1 & = {\textstyle\frac{2\,q_0\, x_1\,\Bigl (1-\sin^2 x_0\, [q_0+(1-q_0)\, (K(q_0)-E(q_0))
]\Bigr ) +(K(q_0)
-E(q_0))\, q_1\, \sin x_0\, \cos x_0}{4\, (E(q_0)-K(q_0))^2\, (q_0-1)\, \sqrt{1-q_0\,\sin^2 x_0}}}\, .
\end{align}
One might wonder whether from the gauge-theory perspective it makes
sense for the splitting parameter $x$ and the outgoing angular
momentum fraction $\beta_{34}$ to be dependent on~${\cal J}$.  After
all, the splitting Hamiltonian commutes with the R-charge
operators~${\cal J}_{12}$ and~${\cal J}_{34}$. Hence, going up higher
in perturbation theory should not induce coupling-constant dependent
modifications to the R-charges of the outgoing strings.  However, the
reason why (\ref{e:b34pert}) is a sensible result is that the
semi-classical string calculation captures only a part (namely the
maximum) of the full quantum surface of the decay process. The
position of the maximum varies as we go higher up in perturbation
theory. At each order in perturbation theory, the most probable
outgoing string with fixed~${\cal J}_{12}^I$ is carrying a different
${\cal J}_{34}^I$. This effectively means that the maximal probability
varies with~${\cal J}$.

\begin{figure}[t]
\begin{center}
\includegraphics*[width=.5\textwidth]{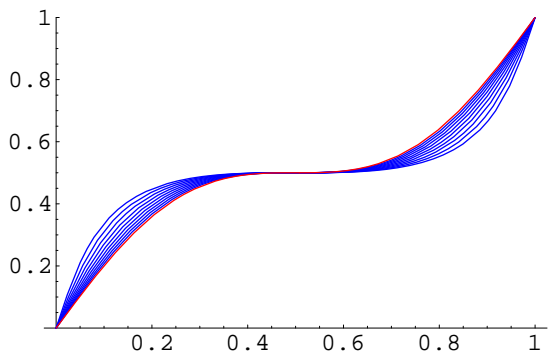}
\caption{Plot of the relation between $\beta_{12}$ (horizontal) and
  $\beta^0_{34}$ (vertical) as defined
  in~\protect\eqn{e:beta1234def}. The various curves correspond to
  various values for the filling fraction $\alpha\in[0.05,\ldots
  0.5]$.}
\label{f:spins-filling}
\end{center}
\end{figure}

In figure~\ref{f:spins-filling} we plot a collection of functions
$\beta_{34}^0(\beta_{12})$ for values of the filling fraction before
the split ranging from~$0.05$ to~$0.5$. These are all computed using
the leading values for the parameter~$q_0$. We choose to restrict to
this region of the filling fraction since the folded string
solution~(\ref{e:FT2}) does not posses a ${\cal J}_{12}
\leftrightarrow {\cal J}_{34}$ symmetry\footnote{On the other hand the
sigma model is, as expected, invariant under this symmetry. The
implementation of this symmetry requires the simultaneous
transformations ${\cal \phi}_1 \leftrightarrow {\cal \phi}_2$, and
$\psi \leftrightarrow {\pi \over 2} - \psi$.} and only the solutions
with $\alpha<0.5$ have been identified on the gauge
side~\cite{Beisert:2003ea}. The solutions with $\alpha>0.5$ are
conjectured to correspond to operators of higher bare dimension, which
do not have a BMN limit.  Note that figure~\ref{f:spins-filling}
possesses a symmetry with respect to the point $(0.5,0.5)$ as a
consequence of the geometry of the folded string ($\psi(\sigma)=
\psi(\pi- \sigma)$). Note also that the point $(1,1)$ corresponds to
the unsplit string.

The second functional relation we want to obtain is a relation between
the energy of the first outgoing piece~${\cal E}^I$ and the parameters $({\cal
J}, \alpha, \beta_{12})$. Eliminating $\kappa, a, \omega_1$ and~$\omega_2$ from 
equation (\ref{e:qsplit}) using (\ref{e:enIII}), (\ref{e:K+F}) and (\ref{e:J12I}) 
leaves us with
\begin{equation}
\label{e:qeqI}
\frac{q}{\pi^2} = \frac{({\cal E}^I)^2}{(K(q)+F(x;q))^2} -
\frac{({\cal J}_{12}^I)^2}{(E(q)+E(x;q))^2} \, ,
\end{equation}
which is the split analogue of \eqn{e:qeq}. The
``mirror'' equation for the second half of the string is obtained from
\eqn{e:qeqI} by replacing the indices $I\to II$ and $x\to -x$. Using
equations~\eqn{eq:beta-exp1} and~\eqn{eq:beta-exp2}, this can be simplified to
\begin{equation}
\label{e:P2}
\frac{4}{\pi^2}\, q  = \frac{({\cal E}^I)^2}{[\beta_{34}\,K(q)+(\beta_{12}-\beta_{34})\, E(q)]^2} 
- \frac{({\cal J}_{12})^2}{E(q)^2} \,.
\end{equation}
Combining this with \eqn{e:qeq} we learn that
\begin{equation}
\label{e:EIfinal}
{\cal E}^I=\Bigl(\beta_{34}+(\beta_{12}-\beta_{34})\, \frac{E(q)}{K(q)}\Bigr)\, {\cal E}\, ,
\qquad {\cal E}^{II}= {\cal E} - {\cal E}^I\, .
\end{equation}
This equation, together with equation~\eqn{eq:beta-exp2} for
$\beta_{34}$ and equation~\eqn{qexp} for $q$, defines ${\cal E}^{I/II}({\cal
J}, \alpha, \beta_{12})$ as a series in $1/{\cal J }$,
\begin{equation}
\label{e:EI/IIexp}
{\cal E}^{I/II}= {\cal J}\,{\cal E}^{I/II}_0 +  {\cal E}^{I/II}_1\, \frac{1}{{\cal J}}  + \ldots \, .
\end{equation}
The first coefficient in the expansion is given by
\begin{equation}
{\cal E}^{I}_0 = \frac{{\cal J}^I_{12}}{{\cal J}} + \frac{{\cal J}^{I,0}_{34}}{{\cal J}} = (1- \alpha)\, \beta_{12} + \alpha\, \beta_{34}^0 \, , 
\end{equation}
and is in agreement with
the (trivial) gauge theory prediction: the two decay products (single trace operators)
have engineering dimensions $J_{12}^I$ and $J_{34}^{I,0}$.
\begin{figure}[t]
\begin{center}
\includegraphics*[width=.5\textwidth]{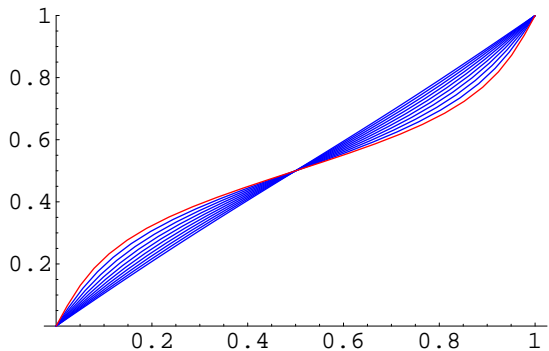}
\end{center}
\caption{The energy ${\cal E}_0^I$ of the first outgoing string as a
  function of $\beta_{12}$, for various filling fractions
  $\alpha\in[0.05,\ldots,0.5]$. The straight line corresponds to
  $\alpha=0.5$.}
\label{energy-spin}
\end{figure}
In figure~\ref{energy-spin} we plot the energy of the first string
piece as a function of $\beta_{12}$, for various filling
fractions. The coefficient at order $1/{\cal J}$ of \eqn{e:EI/IIexp}
reads
\begin{equation}
{\cal E}^{I}_1 ={\textstyle \frac{2}{\pi^2}}\, K(q_0)\, \Bigl [ (K(q_0)-E(q_0))\, \beta_{34}^0\, (q_0-1) + \beta_{12}\,
E(q_0)\, q_0 \Bigr ] + \beta_{34}^1\, (1-{\textstyle \frac{E(q_0)}{K(q_0)}}) \, ,
\end{equation}
which yields a prediction of the anomalous dimension at one loop of the first 
decay product (single trace operator)  in the dual gauge theory.

\begin{figure}[t]
\begin{center}
\includegraphics*[width=.5\textwidth]{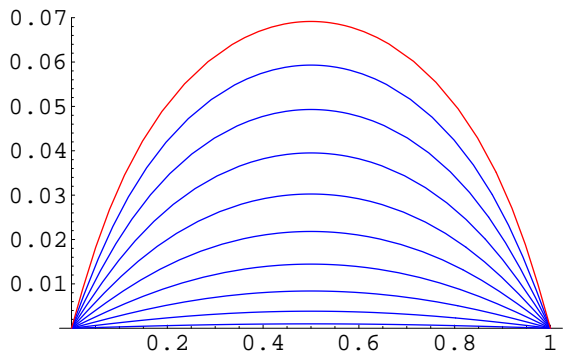}
\end{center}
\caption{The combination of new angular charges $\Delta:= ({\cal
J}_{14}^{I})^2 - ({\cal J}_{23}^{I})^2$ plotted as a function of
  $\beta_{12}$, again for $\alpha$ in the range $[0.05,\ldots,0.5]$. 
 The upper curve corresponds to $\alpha=0.5$.}
\label{f:Delta}
\end{figure}

The third functional relation is obtained by eliminating~$x$
from~\eqn{eq:beta-exp1} and~\eqn{e:music}, after which one can express
$\Delta:= ({\cal J}_{14}^{I})^2 - ({\cal J}_{23}^{I})^2$ 
as a function of $\beta_{12}$ in an expansion in $1/{\cal
  J}$. The corresponding plot is given in figure~\ref{f:Delta}.

\subsection{Higher charges and traces of integrability in the splitting process}

Thus far we have only discussed the behavior of the string energy and
angular momenta under the decay process. However, the classical string
sigma model is known to possess an infinite number of local, conserved
and commuting charges $Q_n$ due to its integrability
\cite{Pohlmeyer:1975nb,Ogielski:1979hv,DeVega:1992xc}, the first
non-vanishing of which is the Hamiltonian~$Q_2={\cal H}$.  These were
written down explicitly in the work of~\cite{Arutyunov:2003rg} for the
folded string solution in terms of a generating functional.  On the
other hand, one does not expect the string sigma model to remain
integrable once string interactions are included (i.e.~when $g_s\neq
0$). This may be seen explicitly from the dual gauge theory side:
nonplanar graphs break the integrability of the planar
theory\footnote{Concretely, one observes degeneracies in the spectrum
of planar~${\cal N}\!=\!4$ super-Yang-Mills (the so-called ``planar
parity pairs'') which may be attributed to the existence of a higher
conserved charge~\cite{Beisert:2003tq}. These degeneracies are lifted,
however, by~$1/N$ corrections: a clear signal of the breakdown of
integrability.}.  Nevertheless it is obvious that, for the
semi-classical decay process we are studying here, the higher
charges~$Q_n$ {\it are} conserved. This conservation follows from the
same logic that was used for the calculation of the energy and angular
momenta. If the initial charges are given via a charge density
as~$Q_n=\int\!{\rm d}\sigma\, q_n(\sigma,\tau)$, then the charges of
the outgoing strings after the split are simply
\begin{equation}
\label{e:Qnsplit}
Q_n^{I} = \int_0^{2 \pi a}\!{\rm d}\sigma\, q_n(\sigma, \tau) \, , \quad Q_{n}^{II} =
Q_n - Q^{I} \, .
\end{equation}
Here one uses the charge densities $q_n(\sigma, \tau)$ {\it before}
the split. In appendix~A we explicitly derive the generating
functional for the commuting charges of the outgoing strings by
generalizing the work of~\cite{Arutyunov:2003rg}. As a side remark let
us note that this knowledge could in principle be used to construct
the explicit form of the outgoing string solutions~$X^{I}(\tau,\sigma)$
and $X^{II}(\tau,\sigma)$.

How is this result to be reconciled with the breakdown of integrability 
at~$g_s\neq 0$? Again we need to remember that the quantum string decay 
leads to a full surface of possible decay channels, which generically will
not preserve the charges beyond~$Q_2$. A subset of channels will, however,
preserve all~$Q_n$. It is precisely this subsector which should capture
the semiclassical string decay analyzed in the previous subsections
and is expected to dominate the decay amplitude.

\section{The decay from the gauge theory side}
\label{s:gauge_theory}

Let us now turn to the discussion of the splitting process in the dual
gauge theory. Our exposition will not be complete, but we will set the
scene for the full calculation and point out the technical
difficulties that one will have to face.

In the large-$N$ limit, the dilatation operator of ${\cal N}=4$
super-Yang-Mills factorizes as the product of a universal space-time
dependent factor times a combinatorial factor acting on the fields
inside composite operators. The string splitting vertex is encoded in
the non-planar piece of this dilatation operator. In the
relevant~SU(2) sector of two chiral complex scalar adjoint fields~$Z$
and~$W$ the (space-time independent part of the) dilatation operator
is known to be~\cite{Beisert:2002bb}
\begin{equation}
D_2=-\frac{g_{\rm YM}^2}{8\pi^2}\, \Tr [Z,W][\check Z, \check W] \, ,
\end{equation}
where $\check Z_{ab}:=\delta/\delta Z_{ba}$ is the matrix derivative
(for a pedagogical derivation see \cite{Plefka:2003nb}).
The action of this operator can be expressed in the language of spin
chains, by considering the action of $D_2$ on two fields in an
arbitrary single trace operator $\Tr(W A Z B)$. One finds
\begin{multline}
D_2\circ \Tr(W A Z B) =\\
 \frac{g_{\rm YM}^2}{8\pi^2}\, \Tr A\, \Bigl ( \Tr(W Z B) -\Tr(Z W B) \Bigr )
+ \frac{g_{\rm YM}^2}{8\pi^2}\, \Tr B\, \Bigl ( \Tr(Z W A) -\Tr(W Z A) \Bigr )\,.
\end{multline}
The planar (nearest neighbor) contribution is obtained when~$A$ is the
identity operator:
\begin{equation}
D_2^{\rm planar} = \frac{g_{\rm YM}^2\, N}{8\pi^2} \sum_{i=1}^L (\delta_{i,i+1} - P_{i,i+1})\,,
\end{equation}
with $P_{i,j}$ the permutation operator permuting the fields (spins)
at sites $i$ and $j$. The system described by $D^{\text{planar}}_2$ is
nothing but the Heisenberg XXX${}_{1/2}$ model
\cite{Minahan:2002ve}. The non-planar contribution may be written as
\begin{equation}
\label{e:D2splitting}
D_2^{\rm splitting} = \frac{g_{\rm YM}^2}{8\pi^2} \sum_{i,j}
(\delta_{i,j} - P_{i,j})\, {\cal S}_{ij}\,,
\end{equation}
where the sum is over non-nearest neighbors and the splitting operator acts on
sites~$i$ and~$j$ as ($\phi_k\in {Z,W}$)
\begin{multline}
{\cal S}_{ij} \circ \Tr(\phi_1\ldots \phi_{i-1} \phi_i \phi_{i+1} \ldots \phi_{j-1} \phi_j \phi_{j+1} \ldots
\phi_L) = \\[1ex]
\Tr(\phi_1\ldots \phi_i \phi_j \ldots \phi_L)\, 
\Tr(\phi_{i+1} \ldots \phi_{j-1})  + \Tr(\phi_1\ldots \phi_{i-1} \phi_{j+1} \ldots \phi_L)\, 
\Tr(\phi_i \ldots \phi_{j})  \, .
\end{multline}
That is, we have a Heisenberg exchange interaction multiplied by a
chain splitting operation (see also \cite{Bellucci:2004ru} for a
related discussion).

While the dilatation operator is thus under control, the initial gauge
theory operator dual to the single folded string solution with angular
momenta ${\cal J}_{12}$ and ${\cal J}_{34}$ is less understood. The
dual gauge operator may be written as
\begin{equation}
\label{e:sketchstate}
\Tr(Z^{{\cal J}_{12}}W^{{\cal J}_{34}}) + \ldots
\end{equation}
where the dots stand for suitable permutations of the $Z$ and $W$'s --
which are of essential importance for the evaluation of decay amplitudes!
The spin chain picture has proved to be very
efficient for the task of diagonalizing  $D_2^{\rm planar}$ for long
operators (${\cal J}\to\infty)$ with the technology of the Bethe ansatz.
There every eigenstate of the ``free'' Hamiltonian $D_2^{\rm planar}$
is parametrized by a set of Bethe roots $\lambda_i$ with
$i=1,\ldots,{\cal J}_{34}$,
which are determined through the Bethe equations
\begin{equation}
\label{e:BetheEq}
\left ( \frac{\lambda_i+i/2}{\lambda_i-i/2}\right )^L=\prod_{k\neq i}
\left ( \frac{\lambda_i-\lambda_k+i}{\lambda_i-\lambda_k-i}\right ) \,   ,
\qquad
\prod_{i=1}^{J_{34}}\frac{\lambda_i+i/2}{\lambda_i-i/2}=1\, ,
\end{equation}
where $L:={\cal J}_{12}+{\cal J}_{34}={\cal J}$.  The corresponding
eigenstate may then be written down explicitly as follows. Denote by
$|\{m_1,m_2,\ldots, m_{J_{34}}\}\rangle_{L}$ the single trace operator
of length $L$ with $W$'s appearing at positions $m_i$, e.g.
$$
|\{ 1,3,4\}\rangle_{L=7} = \Tr(W ZWWZZZ)\, .
$$
Introduce the quasi-momenta $p_i$ and the scattering phases $\varphi_{ij}$
\begin{equation}
p_i:= -i \, \ln\left ( \frac{\lambda_i+i/2}{\lambda_i-i/2}\right )\, ,
\qquad
\varphi_{i,j}:= -i\, \ln \left( \frac{\lambda_i-\lambda_k+i}{\lambda_i-\lambda_k-i}
\right )\, ,
\end{equation}
then the eigenstate of  $D_2^{\rm planar}$ may be written down
explicitly \cite{Bethe:1931hc}\footnote{For a nice hands-on review of this topic
see \cite{Karbach:1997}.}. It is the rather formidable object 
\begin{align}
\label{e:BetheState}
|\psi\rangle =\qquad\qquad &\\ \sum_{\substack{
1\leq m_1<m_2<\ldots \\[1ex]
\ldots <m_{{\cal J}_{34}} \leq L}}&\;\;
\sum_{{\cal P}\in {\rm Perm}_{{\cal J}_{34}}} \exp\Bigl [{i\sum_{i=1}^{{\cal J}_{34}} p_{{\cal P}(i)}
\cdot m_i + \frac{i}{2}\sum_{i<j}^{{\cal J}_{34}}\varphi_{{\cal P}(i),{\cal P}(j)}}\Bigr ]\,
\Bigl |\{m_1,m_2,\ldots,m_{{\cal J}_{34}}\}\Bigr \rangle_{L} \nonumber
\end{align}
where the second sum is over all ${\cal J}_{34}!$ permutations of the
labels $\{1,2,3,\ldots, {\cal J}_{34}\}$.  As Bethe showed in 1931 this is an
eigenstate of the free Hamiltonian
\begin{equation}
\label{e:BetheEnergy}
D_2^{\rm planar}\, |\psi\rangle = 
\frac{g_{\rm YM}^2\, N}{2\pi^2} \sum_{i=1}^{{\cal J}_{34}} \sin^2\left(\frac{p_i}{2}\right)\,
|\psi\rangle \, .
\end{equation}
In order to make contact to our semiclassical string considerations we
need to take the thermodynamic limit $L,{\cal J}_{34}\to \infty$ with
${\cal J}_{34}/L=\alpha$ fixed. Due to the unknown structure of the
continuum limit of the permutation group the Bethe wave function
$|\psi\rangle$ becomes a monstrous object in this limit\footnote{If
one stays with a small number of impurities ${\cal J}_{34}$ and takes
$L\to\infty$ the state remains manageable and is dual to excited
states of the plane wave superstring in the BMN correspondence
\cite{Berenstein:2002jq}.  In this limit one finds $p_i=n_i/{\cal
{\cal J}}$ with integer $n_i$ and $\varphi_{i,j}\to 0$.}.  This is in
stark contrast to the Bethe equations, which actually simplify in the
same limit.  Even worse, we would now want to act with the splitting
Hamiltonian $D_2^{\text{splitting}}$ of \eqn{e:D2splitting}
on~$|\psi\rangle$ in the thermodynamic limit, in order to describe the
quantum decay of the semiclassical folded string solution.  This is
the core of the problem which hampers a direct analytic computation of
the splitting in the gauge theory.  In principle one could attempt to
address this problem numerically. Here however, one faces technical
limitations, as the minimal length of the spin chain for which
distinguishable structures limiting to the continuum folded string
configuration start to emerge is~26 (with half filling fraction)
\cite{Beisert:2003xu}.  The corresponding wave function~$|\psi\rangle$
contains roughly~$4\cdot 10^5$ terms, many of which have
coefficients of the same order.

An alert reader might wonder whether the gauge theory again develops
an effective genus counting parameter ${\cal J}^2/N$ in the
thermodynamic limit (${\cal J}_{12}$, ${\cal J}_{34}$, $N\to \infty$)
as it does in the BMN limit where ${\cal J}_{34}$ remains finite
\cite{Kristjansen:2002bb,Constable:2002hw}.  It is very plausible that
this is the case. Indeed a simple pilot computation of the free theory
two point function of two operators of type~\eqn{e:sketchstate} in the
this limit confirms the expectation.  One finds using the method of
``highways''~\cite{Kristjansen:2002bb} up to genus one
\begin{align}
\langle \Tr(\bar Z^J\, \bar W^J)\, \Tr (Z^J\, W^J)\rangle &=
N^{2J}\, \Bigl ( 1 +\frac{1}{N^2}\, \Bigl ( 2\,\Bigl  [ \, \left (
\begin{matrix} J+1\\3\end{matrix}\right )+\left (
\begin{matrix}J+1\\4\end{matrix}\right )\, \Bigr] + (J-1)^2\, \Bigr )+\ldots \Bigr )
\nonumber\\
&\to\, N^{2J}\, \Bigl( 1 +12\,  \frac{J^4}{N^2} +\ldots \Bigr )
\end{align}
which displays the expected $J^4/N^2$ scaling behavior.

It is instructive to look at the decay of a number of lower eigenstates 
of $D_2^{\rm planar}$ as toy calculations exemplifying the general
logic of the quantum decay. 
All eigenstates $D_2^{\rm planar}$ may be classified by Bethe roots,
or equivalently by the values of the local, conserved charges
$Q_i$. At one loop, these are given by the moments of the resolvent,
and the first nonvanishing charge corresponds to the one loop anomalous
dimension of the state $E=Q_2$. On top of this, all states carry a
representation of the global R-symmetry group. In the~SU(2) sector,
these are realized through the operators
\begin{equation}
\label{su(2)op}
J_z \equiv J_{12 } - J_{34} =\Tr(W\check W- Z\check Z)\, , \qquad J_+=\Tr(W\check Z)\, ,\qquad J_-=\Tr(Z\check W)\, ,
\end{equation}
where, geometrically, the operators $J_{12}$ and $J_{34}$ correspond
to the rotations in the two two-planes $W, \bar{W}$ and $Z,
\bar{Z}$. It is easy to check that the operators (\ref{su(2)op}) obey
an su(2) algebra: $[J_+,J_-]=J_z$ and $[J_z,J_\pm]=\pm 2\, J_\pm$. The
full (planar and non-planar) dilatation operator $D_2$ indeed commutes
with these operators: $[D_2,J_z]=0=[D_2,J_\pm]$. Hence, the total spin
and the $J_z$ charge of a given initial state is conserved in the
decay process.  Highest-weight single trace states obey $J_-\, |{\rm
HWS}\rangle =0$ and correspond to ensembles of Bethe roots at finite
values. Acting with $J_+$ on $|{\rm HWS}\rangle$ increases the number
of impurities but leaves the energy invariant. This corresponds to
adding Bethe roots at infinity. Note also that expectation values of
(``non-Cartan'') operators $J_x$ and $J_y$ in the $|{\rm HWS} \rangle$
obviously vanish.

All the local conserved charges of the Heisenberg XXX${}_{1/2}$ are
known explicitly \cite{Grabowski:1994ae}. As we illustrate now, these are generically not
conserved in the decay process. Explicitly, the first three charges are
given by~\cite{Beisert:2003tq}
\begin{equation}
\begin{aligned}
Q_2 &= 2\sum_{i=1}^L ( 1 -P_{i,i+1} ) \,,\\
Q_3  &= 4\sum_{i=1}^L (P_{i,i+1}\, P_{i+1,i+2} - P_{i+1,i+2}\, P_{i,i+1} ) \,,\\
Q_4 &=\sum_{i=1}^L (
\begin{aligned}[t]
&  -2 P_{i,i+1} + P_{i,i+1}\, P_{i+1,i+2} + P_{i+1,i+2}\, P_{i,i+1}\\
&  +P_{i,i+1}\, P_{i+2,i+3}\, P_{i+1,i+2} + P_{i+1,i+2}\, P_{i,i+1}\, P_{i+2,i+3} \\
& - P_{i,i+1}\, P_{i+1,i+2}\, P_{i+2,i+3}- P_{i+2,i+3}\, P_{i+1,i+2}\, P_{i,i+1})\,.
\end{aligned}
\end{aligned}
\end{equation}
The odd charges have negative parity and either annihilate a
highest-weight state, or pair them to a partner of opposite parity,
which is degenerate in energy.  Let us for example, look at the decay
of two highest-weight states of length 9,
\begin{align}
{\cal O}_-^{9,4}=& - \Tr(Z^4WZW^3)+\Tr(Z^4W^3ZW)+\Tr(Z^3WZ^2W^3)-\Tr(Z^3W^3Z^2W)
\nonumber\\\label{e:O9s}
&+\Tr(Z^3WZWZW^2)
-\Tr(Z^3W^2ZWZW)  \qquad q_2=5,\, q_4=1\nonumber\\
{\cal O}_+^{9,2}=&\, \Tr(Z^7W^2)-\Tr(Z^6WZW) \qquad  \qquad\qquad\quad \qquad q_2=4,\, q_4=-16\, ,
\end{align}
where we have also spelled out their charges with respect to the~$Q_2$
and~$Q_4$ operators. The first state decays into $Q_4$ non-conserving
constituents:
\begin{align}
D_2^{\rm splitting}\circ {\cal O}_-^{9,4} = 12\Bigl (\,  \Tr(Z^2)\, {\cal O}^{7,4}_- + \Tr(ZW)
\, {\cal O}_-^{7,3}\, \Bigr )\,,
\end{align}
where 
\begin{equation}
\begin{aligned}
{\cal O}^{7,4}_- &= \Tr(Z^2W^3ZW)-\Tr(Z^2WZW^3)\,,\qquad q_2=5, q_4= -5 \,, \\
{\cal O}^{7,3}_- &= \Tr(Z^3WZW^2)-\Tr(Z^3W^2ZW)\,,\qquad q_2=5, q_4= -5 \,.
\end{aligned}
\end{equation}
The protected states $\Tr(Z^2)$ and $\Tr(ZW)$ have vanishing charges. Therefore $Q_4$
is not conserved in both decay channels. Note also that
\begin{equation}
J_-\circ {\cal O}_-^{7,4}=-{\cal O}_-^{7,3}\quad J_-\circ \Tr(ZW)=\Tr(Z^2) \quad
J_-\circ \{ \, {\cal O}_-^{9,4}, {\cal O}_-^{7,3} , \Tr(Z^2)\, \}=0\, .
\end{equation}
Hence a highest-weight state does not necessarily decay into products
of highest-weight states. 

The highest-weight state ${\cal O}^{9,2}_+$
of~\eqn{e:O9s} on the other hand has only a single decay channel
\begin{equation}
D_2^{\rm splitting}\circ {\cal O}_+^{9,2} = 8\, \Tr(Z^2)\, {\cal O}_+^{5,2} +
(\mbox{non $Q_2$ preserving channels})\,,
\end{equation}
where
\begin{equation}
{\cal O}_+^{5,2} = \Tr(Z^3W^2)-\Tr(Z^2WZW)\,, \qquad q_2=4,\, q_4=-16
\end{equation}
Hence here the higher local charge $Q_4$ is conserved. 

In summary, by considering the decays of short operators, we see that
the highest-weight states do not need to decay into a product of
highest-weight states, and that higher charges are not preserved in
this decay process. However, in the thermodynamic limit we expect the
decay to be dominated by the channels which \emph{do} preserve all
higher charges. The outgoing states are not highest-weight
states. This can be seen from the fact that in the semiclassical
calculations, equations~(\ref{e:Jnew1}) and (\ref{e:Jnew2}) indicate
that the expectation values for the ``non-Cartan'' angular momenta,
$\langle J_{13} \rangle$ and $\langle J_{24} \rangle$, are
non-vanishing after the decay.
\medskip

\section{Outlook}

The folded spinning string for which we have calculated the classical
decay process has been identified on the gauge theory
side~\cite{Beisert:2003xu} by solving the Bethe
equations~\eqn{e:BetheEq}. In the sector in which the number of
impurities is odd, it appears as the first highest-weight state above
the vacuum. Using~\eqn{e:BetheEnergy} it is possible to compute its
energy directly from the Bethe roots. For half-filling, an analysis of
various spin chain lengths (up to length~46) has shown that this
energy is approximated by
\begin{equation}
{\cal E} - {\cal J}{\cal E}_0 = \frac{0.356}{{\cal J}} + \ldots\,.
\end{equation}
This matches the result computed from~\eqn{e:q0} and~\eqn{e:E0E1} for
$\alpha=1/2$.

In principle, one can apply the machinery of
section~\ref{s:gauge_theory} in order to study the decay of this
state. The charges which enter in the classical decay relations on the
string side, as depicted in
figure~\ref{f:spins-filling}--\ref{f:Delta}, have direct analogues in
the spin chain. As an example, the parameters~$\beta_{12}$ and
$\beta_{34}$ are related to filling fractions~$\alpha$ and~$\alpha^I$
and chain lengths~$L$ and~$L^I$ according to
\begin{equation}
\beta_{34} \leftrightarrow \frac{\alpha^I L^I}{\alpha L}\,,\quad
\beta_{12} \leftrightarrow \frac{(1-\alpha^I)L^I}{(1-\alpha)L}\,.
\end{equation}
As explained in the introduction, the splitting amplitude of the spin
chain is expected to attain its maximum over the semi-classical curves
found in section~\ref{s:relations}.

However, as we have already alluded to in the previous section, the
main obstacle is the complexity of the Bethe wave
function~\eqn{e:BetheState}. Even for moderately large spin chains,
the Bethe state is a monstrous object, which complicates a brute force
analysis through a numerical treatment. One possible simplification
can perhaps be obtained by using the coherent state wave function
of~\cite{Kruczenski:2003gt}. However, a potential problem in this
approach seems to arise from the inability to write down wave
functions for the outgoing strings.

An additional guideline for a better analytic understanding is the
existence of the higher local charges.  The decay channels in which
these charges are conserved are expected to correspond to
semi-classical decay, and form only a small subsector of all possible
channels. We will return to this spin chain analysis in future work.

Finally an interesting question which deserves investigation 
concerns the circular string solution of Frolov and
Tseytlin~\cite{Frolov:2003qc}. This solution has also been
successfully matched to gauge theory \cite{Beisert:2003xu}. Clearly
the circular string is semiclasssically stable, as it does not self
intersect. How does this property reflect itself in the dual gauge or
spin chain description?

\acknowledgments

We wish to thank N. Beisert, M. Staudacher, K. Zarembo
and in particular G. Arutyunov for useful discussions and
comments.

\newpage
\appendix
\section{Generating function for commuting charges of the outgoing string}

In the case of the~$O(6)$ model, an infinite set of commuting charges
can be constructed using the so-called B\"acklund
transformations. Namely, given a particular solution $X^{\mu}_{0}$, by
solving a set of equations~\cite{Ogielski:1979hv} one can derive a ``dressed''
solution $X^{\mu}(\gamma)$ (with $X^{\mu}(0)= X^{\mu}_0$) as a power
series in the spectral parameter~$\gamma$. The coefficients in the
series are determined though certain recursive relations.
The generating function for the charges is then obtained from the dressed
solution as
\begin{equation}
\label{gene}
\Gamma (\gamma) = \int\! {{\rm d} \sigma \over 4 \pi} \left(  
\gamma (X(\gamma) \cdot X_{\xi}) + \gamma^3 (X(\gamma) \cdot X_{\eta}) \right) \, ,
\end{equation}
where $\xi=\frac{1}{2}(\tau+\sigma)$ and $\eta=\frac{1}{2}(\tau-\sigma)$,
while subscripts denote partial derivatives.  In the case of the folded string,
a solution to the B\"acklund transformation has been constructed
exactly (to all orders in $\gamma$) in~\cite{Arutyunov:2003rg} and is
given by
\begin{equation}
\label{solu}
Z_i(\gamma) \equiv X_i + iX_{i+3}= r_i(\sigma, \gamma) e^{i \alpha_i(\sigma, \gamma)}
e^{i \omega_i \tau} \, ,  \,  \quad (i=1,2,3)\,.
\end{equation}
The~$r_i$ are defined as
\begin{equation}
r_1(\sigma,\gamma) = \dn(\sqrt{\omega_{21}^2} \sigma+ \nu,t) \, , \quad r_2(\sigma,\gamma)
= \sqrt{t} \sn(\sqrt{\omega_{21}^2} \sigma+ \nu,t) \, , \quad r_3(\sigma,\gamma) =0 \, .
\end{equation}
The constant phases~$\alpha_i$ are given by
\begin{equation}
\cos \alpha_1 = {1- \gamma^2 \over 1 + \gamma^2}{1\over \dn \nu} \, ,
\quad \cos \alpha_2 = {1- \gamma^2 \over 1 + \gamma^2} {\cn \nu \over
  \dn \nu} \,.
\end{equation}
The functional dependence of the parameter~$\nu$ on the spectral
parameter~$\gamma$ is given by the equation
\begin{equation}
1- {\omega_1^2 \over \omega_{21}^2}{\sn^2 \nu \over \cn^2 \nu} - \left(  {1 - \gamma^2 \over 1 + \gamma^2} \right)^2 {1\over 1 - t^2 \sn^2 \nu} = 0 \, . 
\end{equation}
In addition, the periodicity condition for the folded string implies
the relation
\begin{equation}
{\pi \over 2} \sqrt{\omega_{21}^2} = K(t) \, , \quad \omega_{21}^2 \equiv \omega_2^2 - \omega_1^2 \, ,
\end{equation}
which is~\eqn{e:omK} of the main text.

Inserting~\eqn{solu} into the expression for the generating function
of the charges~\eqn{gene}, one obtains
\begin{equation}
\Gamma(\gamma;a) = \gamma \int_0^{2 \pi a} {{\rm d}\sigma \over 2 \pi}
r_i(\sigma,\gamma)\,\big[(1-\gamma^2) \cos \alpha_i\, r_i'(\sigma,0) + (1+ 
\gamma^2)\omega_i\, \sin \alpha_i\, r_i(\sigma,0) \big] \, ,
\end{equation}
where here the constant~$a$ is the splitting parameter. For $a=1$, this
integral determines the charges of the incoming string, and it has been
computed in~\cite{Arutyunov:2003rg}.  To evaluate this integral for a
generic value of the parameter $a$, and hence determine the generating
functions for commuting charges of the outgoing strings, one needs the
following four integrals:
\begin{multline}
\int_0^{2 \pi a} {{\rm d} \sigma \over 4 \pi} r_1(\sigma,\gamma)\, r_1(\sigma,0) =  {\dn \nu
  \over 2 K(t) \sn^2 \nu} \Big[a K(t) - \cn \nu\, \Pi(a,t \sn^2\nu, t)\Big]
  \\
 + {1\over 16 K(t)}{\cn \nu \over \sn \nu } \ln |A(\nu,t)| \,,
\end{multline}
\begin{multline}
\int_0^{2 \pi a} {{\rm d} \sigma \over 4 \pi} r_2(\sigma,\gamma)\,
  r_2(\sigma, 0) = - {\cn \nu
  \dn \nu
  \over 2 K(t) \sn^2 \nu} \Big[ a K(t) - \Pi(a,t \sn^2\nu, t)\Big] \\-
  {1\over 16 K(t)}{1\over \sn \nu} \ln |A(\nu,t)| \,,
\end{multline}
\begin{multline}
\int_0^{2 \pi a} {{\rm d} \sigma \over 4 \pi} r_1(\sigma,\gamma)\, r_1'(\sigma,0) = {\cn \nu
  \over \pi \sn^3 \nu} \Big[ \dn^2 \nu (K(t)a  -  \Pi(a,t \sn^2\nu, t)) \\
+ \sn^2\nu\, (K(t)a - \tfrac{1}{4} E(\tilde{a},t))\Big] 
- {1\over 8 \pi} { \dn \nu \over \sn^2 \nu} \ln |A(\nu,t)| \,,
\end{multline}
\begin{multline}
\int_0^{2 \pi a} {{\rm d} \sigma \over 4 \pi} r_2(\sigma,\gamma)\, r_2'(\sigma,0) = {1
  \over \pi \sn^3 \nu} \Big[ \cn^2 \nu \dn^2 \nu \Pi(a,t \sn^2, t)   + \sn^2
  \nu\, \tfrac{1}{4} E(\tilde{a},t) \\ - \dn^2 \nu K(t) a \Big] 
- {1\over 8 \pi} { \dn \nu  \cn \nu \over \sn^2 \nu} \ln |A(\nu,t)|
  \,,
\end{multline}
where
\begin{equation}
\label{e:Adef}
A(\nu,t) = t \sn^2 \nu \sn^2\big(4 K(t) a\big) - 1 \qquad\text{and}\qquad
\tilde{a} = \text{am}\big(4 a K(t),  t\big)\,.
\end{equation}
Here $\Pi(a,m^2,t)$ is the incomplete elliptic integral of the third kind, defined as
\begin{equation}
\Pi(a,m^2,t) = {K(t) \over 2 \pi} \int_0^{2 \pi a } {{\rm d} \sigma \over 1- m^2 \sn^2 \left(\frac{2}{\pi} K(t) \sigma,t\right)} \, .
\end{equation}

The expression for the generating function for the charges of the
outgoing string of ``length''-$a$ can be now be written as
\begin{equation}
\begin{aligned}
\Gamma(\gamma;a) & = \Gamma(\gamma)^{I} + \Gamma^{II}(\gamma)\,, \\[1ex]
\Gamma^{I}(\gamma;a) & = - {\gamma \cn \gamma \over (1 + \gamma^2) \pi \dn \nu \sn^3 \nu } \left( B(\nu,\gamma) K(t) + C(\nu,\gamma) \Pi(a,t^2 \sn^2 \nu,t)  \right)\,, \\[1ex]
\Gamma^{II}(\gamma;a) & = {1 \over 4 \pi}{\gamma \over 1 + \gamma^2} {1\over \sn^2 \nu} \left( (1+ \gamma^4) - 2 \cn^2 \nu \gamma^2 \right) \ln |A(\nu,t)| \, ,
\end{aligned}
\end{equation}
where the functions $B(\nu,\gamma)$ and $C(\nu,\gamma)$ are given by
\begin{equation}
\begin{aligned}
B(\nu,\gamma) &= (1-\gamma^2)^2 \left( \left( \sn^2 \nu - \dn^2 \nu
\right) a + {\cn^2 \nu \over a} - 1 \right) - (1+ \gamma^2)^2 {\dn^2
  \nu \over a} + 2 (1 + \gamma^4) \dn^2 \nu \,, \\
C(\nu, \gamma) & = (1 -\gamma^2)^2 \left( \cn^2 \nu\, \left({1\over a} -1\right) + {\dn^2 \nu \over a} - \sn^2 \nu \dn^2 \nu \right) - (1 + \gamma^2)^2 \cn^2 \nu \dn^2 \nu \, \, ,\end{aligned}
\end{equation}
and the function $A(\nu,t)$ is given in~\eqn{e:Adef}.  
\newpage

\begingroup\raggedright\endgroup

\end{document}